\documentclass{article} \usepackage{amsmath} \usepackage{amssymb}
\usepackage{epsfig}
\newcommand{\eg}{\textit{e.g. }}
\newcommand{\ie}{\textit{i.e. }}

\newtheorem{condition}{Condition}

\font\openface=msbm10 at10pt

\def\Integers      {{\hbox{\openface Z}}}

\begin{document}
\title {Macroscopic observables and Lorentz violation in discrete quantum gravity} \author{Joe
Henson} \maketitle
Institute for Theoretical
Physics, University of Utrecht, Minnaert Building, Leuvenlaan 4, 3584 CE
Utrecht, The Netherlands.
\newpage
\begin{abstract}

This article concerns the fate of local Lorentz invariance in
quantum gravity, particularly for approaches in which a discrete
structure replaces continuum spacetime.  Some features of standard
quantum mechanics, presented in a sum-over-histories formulation,
are reviewed, and their consequences for such theories are
discussed. It is argued that, if the individual histories of a
theory give bad approximations to \textit{macroscopic} continuum
properties in some frames, then it is inevitable that the theory
violates Lorentz symmetry.

\end{abstract}
\vskip 1cm

\section{Introduction}

Local Lorentz Invariance, or more specifically the question of
whether this physical principle will be maintained or broken at
the Planck scale, has become a much debated topic both in the
theory and phenomenology of quantum gravity.  On the
phenomenological side, constraints on violation of Lorentz
invariance (termed ``Lorentz violation'' below) are becoming ever
more stringent \cite{Amelino-Camelia:2004hm}, and more arguments
have recently appeared suggesting that Planck scale Lorentz
violation is incompatible with current observations, without (at
best) additional fine-tuning being introduced
\cite{Collins:2004bp,Collins:2006bw}.  With this progress, it
becomes increasingly important for the theoretical side to provide
predictions, at least of a heuristic nature: in any particular
theoretical framework, what becomes of Lorentz invariance?  Is it
possible to maintain it, or must it be broken at some scale?

It is well-known that waves travelling on lattices violate Lorentz
invariance, and this is often used as an argument against
fundamental discreteness.  Some Loop Quantum Gravity (LQG) based
arguments have been made to support this expectation
\cite{Gambini:1998it,Sahlmann:2002qj,Sahlmann:2002qk}, and some
further discussion of the methods used is given in
\cite{Bojowald:2004bb}. However, there are conflicting arguments
for local Lorentz invariance in LQG \cite{Rovelli:2002vp}, and
also some \cite{Livine:2004xy} for the ``third way'' of Doubly
Special Relativity (DSR) \cite{Amelino-Camelia:2005ne} in which
the Planck length or Planck energy is introduced as an invariant
scale along with the speed of light, deforming the Lorentz
transformations (an idea that is critised in
\cite{Sudarsky:2005ua}). If the spin-foam quantum gravity program
is to provide a path-integral formulation of LQG, then it might be
expected that the argument which prevails in the LQG program will
also prevail for these models, and \textit{vice-versa}.

Two interesting strategies have been proposed to evade the
conclusion that Lorentz violation follows from discreteness. One
is to find a form of discreteness that retains Lorentz symmetry at
the level of the continuum approximation.  The causal set
\cite{SpacetimeAsCS,Henson:2006kf} offers such a discretisation of
spacetime, as argued in \cite{Dowker:2003hb} and proved in a
strong sense in \cite{Bombelli:2006}.  This allows waves to travel
on a fixed background causal set without extra Lorentz violating
terms appearing in the dispersion relation \cite{Henson:2006}.
This special property of the causal set is often given as a
motivation for using these structures as the histories in a
quantum gravity sum-over-histories (SOH).

In spin-foam models, this Lorentz invariance of individual
histories has not been claimed, and so a different argument must
be employed.  Since discreteness arises in standard quantum
theories without violating continuum symmetries, it has been
suggested that the same might be true for Lorentz symmetry in
quantum gravity. This line of reasoning has been expanded on in
the context of LQG \cite{Rovelli:2002vp}.  In that article, an
analogy is drawn between the rotation group in standard quantum
mechanics, and the Lorentz group in quantum gravity. It is pointed
out that, although the components of the angular momentum of, say,
an electron, cannot be simultaneously measured, nevertheless any
one can be measured and the theory is still rotationally
invariant. It is claimed that the same reasoning will prevail in
the case of measurements made with respect to different frames in
loop quantum gravity.

If this evasion of Lorentz violation were to be successful in
sum-over-histories approaches such as spin-foams, we would be able
to say that, although the individual histories of our theory
(which are configurations in one Hilbert-space basis, given at all
times) do violate the symmetry, somehow it can still be true that
our measurements do not.  The underlying reasoning is that the
results of measurements need not all directly correspond to
properties of the histories. But in standard quantum theory,
approximate, macroscopic measurements (measurements that we can
make without significantly altering the state of the system)
\textit{do} correspond to properties of the histories. The main
point of this article is that even these macroscopic properties
are put in danger if the histories are not Lorentz invariant.

In the angular momentum example, this danger does not exist, since
all components of the (approximate) angular momentum of a
macroscopic object can be seen as a property of the histories. In
other words, measurements of angular momentum components of, say,
a baseball, can in principle be made in the same basis, up to an
acceptable degree of accuracy.  But there is an important
difference in the case of Lorentz symmetry, more fully explained
later on.  If a history in a quantum gravity SOH was like a
lattice, then even \textit{macroscopic} quantities could fail to
be properties of the history in highly boosted frames. This
situation is qualitatively different from that of rotational
invariance.  It is argued below that this difference will prevent
Lorentz invariance in a class of discrete models.

This article is intended to introduce a different way of looking
at the problem of Lorentz violation in quantum gravity, through
the lens of the sum-over-histories formalism for quantum
mechanics.  As such it is does not take the argument to any level
of technical sophistication, but only presents a framework that,
it is hoped, will facilitate further debate.  Roughly, the
argument proceeds in the following stages:

1) The outcome of the measurement of a macroscopic quantity
corresponds to a property of some subset of the histories in the
SOH.  It follows that, in our quantum gravity SOH, there must be
an approximate correspondence between some of the histories and
Lorentzian manifolds (or at least those properties of Lorentzian
manifolds that we expect to be measurable in principle).  It is
further assumed that some of the histories should have Minkowski
space as an approximation.

2)  A semi-classical state that ``tracks'' Minkowski space is
considered, along with quantities that are defined with respect to
some co-ordinate frame on this Minkowski space.  Consider a
quantity that is macroscopic and measureable in this ``fiducial''
coordinate frame. For Lorentz invariance to hold, this quantity
must be macroscopic and measureable in all frames. Therefore there
must be histories, each of which has properties that correspond to
the outcomes of measurements of the macroscopic quantity in all
frames.

3)  There is a condition on the histories of a quantum gravity SOH
(arguably true for lattices and similar discretisation schemes)
under which no one history can contain such properties in all
frames. It is shown that this leads to either Lorentz violation or
a lack of macroscopically observable properties.

4) Spin foam quantum gravity models are likely to satisfy this
condition if they are discrete at the Planck scale.

The argument relies on the correspondence between the outcomes of
measurements and the properties of histories in the SOH.  This
view, suggested by Feynman in his original paper on the subject
\cite{Feynman:1948ur}, is not currently \textit{en vogue}, and so
a justification of the point is necessary. Because of this, the
first section below is a detour into sum-over-histories quantum
mechanics, which also serves to fix some notation.  As this
viewpoint might be useful for other problems in quantum gravity, a
fuller treatment is given in the appendix.  In subsequent
sections, we return to the specific case of discrete quantum
gravity, covering the other points mentioned above. An outline of
the conclusion is this: Lorentz violation at the level of the
individual histories of a theory, if it causes macroscopic
properties to be badly approximated in some frames, leads to
Lorentz violating predictions.

\section{Properties of histories, outcomes of measurements}

Must there be an approximate correspondence between some of the
histories in our quantum SOH, and the histories of the classical
theory that we wish to approximately recover?  Specifically to
quantum gravity, we might ask: should some of our histories have
Lorentzian manifolds as approximations at some large scale?  The
existence of such a discrete/continuum correspondence is supported
in some of the literature (\eg \cite{SpacetimeAsCS}, and
\cite{Bombelli:2004si} in which such a correspondence is sought
for spatial configurations rather than histories); however, it is
not an uncommon opinion that this will not be necessary (see \eg
\cite{Markopoulou:2002ru}).  The properties of the continuum
manifold need only become evident, it is argued, in quantum sums
over many of the fundamental histories. This view is no doubt
motivated by thinking of properties like the momentum of a quantum
particle. This property is certainly not a property of any history
(if those histories are written in the position basis)
\footnote{A similiar intuition might be noted, coming from the
expansion of path integrals for interacting quantum field theories
written in terms of Feynman diagrams. There, the histories appear
merely as abstract graphs, plus some extra information, that do
not correspond to properties that we eventually measure.  But,
again, the intuition that histories do not correspond to
measurements does not extend to macroscopic properties.}. From
this, it is sometimes concluded that individual histories in the
SOH are of no physical significance.

Here it is argued that this reasoning fails to apply for
properties that are macroscopic in a suitable sense.  The momentum
of an electron may not be a property of a history, but in a
semi-classical state, the momentum of a baseball (measured with
some acceptable degree of accuracy) is. On a suitably
coarse-grained scale, the momentum of the baseball can be read off
from a typical history in the same way as it would have been in
the classical theory. This not only holds in all standard quantum
theories, but is a necessary feature of any physically realistic
quantum theory, as explained in the appendix.

This SOH view of quantum mechanics is nothing but a re-casting of
the standard formalism; it is possible to describe our present,
successful theories in this language.  The following arguments on
Lorentz violation apply only to theories that are compatible with
this standard framework, and so this is introduced as a first
condition for the main argument to hold.

\begin{condition}
\label{c:soh} The quantum gravity theory is compatible with SOH
formalism.
\end{condition}

\subsection{The sum over histories}
\label{s:SOH}

First, we recall some features of standard quantum mechanics.  In
the paper in which he introduced the SOH formalism
\cite{Feynman:1948ur}, Feynman also gave some appealing and useful
ways of picturing it, allowing an easy connection between the
formalism and observations.

The theory has a history space $\Omega$, the elements of which are
the histories.  A history gives a possible configuration of the
system in question (in some special Hilbert-space basis) at all
times under consideration.  For example, in the case of the
Schr\"odinger particle they are continuous (but not necessarily
differentiable) paths $x(t)$ between some initial and final times.
A brief quote from Feynman shows how measurements are dealt with
in this formalism:

\begin{quote}
The probability that a particle will be found to have a path x(t)
lying somewhere within a region of space-time is the square of a
sum of contributions, one from each path in the region.
\cite{Feynman:1948ur}
\end{quote}

So, in measurement situations, our quantum theory associates
probabilities to properties of the histories. When we talk of a
property $X$ of a history, we can associate to it a set of
histories that have that property, $\Gamma(X) \subset \Omega$, and
a question ``Does the system have property $X$?''. For example,
the property may be that $a<x(t)<b$ for some particular $a$, $b$
and $t$. Answering the associated question requires a measurement,
in this case a standard measurement. More generally, we might ask
``did the particle's world-line pass through spacetime region
$R$?'' for some particular $R$, which amounts to a non-standard,
``continuous'' measurement.

If we observe a property $X$ we condition on the set of histories
$\Gamma(X)$ with that property, throwing away all the histories
not in $\Gamma(X)$ and carrying out the appropriate
renormalisation of the probabilities (the equivalent of the
``collapse of the wave function'' from this perspective).  We
might want to carry out another measurement, finding that the
system has another property, $Y$, and we would then further
condition on the smaller set of histories $\Gamma(X) \cap
\Gamma(Y)$, and so on.  We will call this subset of the history
space the \textit{measured set of histories} or \textit{measured
set}. See \cite{Sinha:1991cj} for an application of the formalism
to a familiar experiment.

From the quote, we can see that the individual histories in the
measured set do have at least some significant properties.  The
properties of the system that we observe are those shared by all
the histories in the measured set.  So far, there is no need to
talk about anything but properties of histories in order to
describe the observations\footnote{Note that we have not stepped
outside of the normal interpretation of quantum mechanics.
Physical properties can therefore be interpreted here in an
operationalist sense. However, if preferred, it is also possible
to apply a ``decoherent histories'' style interpretation (see \eg
\cite{Hartle:1992as}) without altering the argument.}. But all
these measurements were in the specially selected basis. So the
question becomes:  how much of the observed data can be described
by measurements in this one basis?

As noted above, not all microscopic variables in a quantum system
are directly represented in the histories.  But it presents no
difficulty to perform measurements of all \textit{macroscopic}
quantities in one basis, when they are suitably defined.  Recall,
for example, the properties of a semi-classical, coherent state
for the Schr\"odinger quantum theory: the product of uncertainties
of position and momentum is minimised, and the quantum state, when
peaked on a certain position and momentum, ``tracks'' the
associated classical solution over time.  An approximate
measurement of position (projecting onto a certain macroscopic
range, say) can be made without significantly disturbing the state
-- indeed, this might be taken as a definition of a macroscopic
measurement.  By appeal to the classical theory, we can now see
that a time-sequence of such position measurements must be enough
to approximately determine the conjugate variable, the momentum.
Thus the (approximate) momentum of a macroscopic object can be a
property of the histories in the measured set, just as the
position is -- even though the ``typical path'' can be highly
fractal on smaller scales.

The argument that this is the case for standard theories, and
moreover must be the case for any realistic quantum theory with a
good semi-classical regime, is left for the appendix.  Also
treated there, in section \ref{a:Feynman}, is a method sketched
by Feynman for indirectly finding the results of all microscopic
measurements in one basis.  The conclusion is that \textit{each
individual history in the measured set must have all of the
properties that we observe}.

\subsection{Effective descriptions}
\label{s:eff}

Sometimes our fundamental histories are not directly described in
terms of the properties that we are familiar with at the
macroscopic level. But our observations may be formalised in terms
of these these effective properties.

For example, if our system is a collection of molecules, we might
make a hydrodynamical approximation.  But this continuum
description is of course not perfect.  Some of the properties of
the hydrodynamical description do not have any physical
significance.  An example is the property that a density
perturbation of amplitude $A$ and wavelength $\lambda$ exists in
some region of the fluid.  This ceases to make sense in the
underlying history when $\lambda$ becomes close to the
intermolecular separation.

If an effective property $X$ does not have any corresponding
property in a fundamental history $h$ (\ie if it is impossible to
tell from $h$ whether or not $X$ has happened) then we must
consider $X$ false for that history, and so $h$ would not be
included in $\Gamma(X)$.  Such properties are termed
``undecidable'' for that history (the converse being
``decidable'').

\section{Lorentz invariance}
 \label{s:LI}

These considerations will now be applied to the problem of Lorentz
invariance in quantum gravity.  But what would Lorentz symmetry
mean in this context?

Here our system is spacetime, and our histories, at least on an
effective level, are Lorentzian manifolds.  It is simplest to
assume that there is a semi-classical state that tracks Minkowski
space, in order to talk about global Lorentz invariance. This
should not be a controversial assumption, as much effort has gone
into attempting to describe such states in the various quantum
gravity programs. To say that this state is Lorentz invariant
means that there should be no way to pick out a preferred
direction in spacetime by any measurement. This immediately tells
us that, if we expect Lorentz invariance, then we expect the
semi-classical behaviour to persist in all frames. But
semi-classical behaviour (and what it means to track a classical
solution) should be defined in terms of some class of observable
quantities that we expect to be able to measure.

\subsection{Macroscopic observables in quantum gravity}

Our measurements might be measurements of time and length in one
frame, made at scales appropriate to the semi-classical state, \eg
the informal, very approximate measurements of space and time that
we make every day.  They could also include measurements of other
fields.  Different observers might set out to measure similar
macroscopic quantities in other frames.

Already, the terms ``macroscopic'', ``semi-classical'' and
``tracking Minkowski'' have been used, but no exact definition has
been given.  What exactly are these scales at which observables
can be considered macroscopic, in quantum gravity?  No detailed
investigation shall be launched into here; it is only necessary to
stress the uncontroversial point that some familiar observables
must be considered to be in this class.  It is trivial to observe
good approximations to classical predictions, for example, in
electromagnetic waves moving on an approximately flat background,
for certain values of frequency and amplitude.  Our theory of
quantum gravity must be able to reproduce these results and others
like them, and so measurements of such waves must be in the class
of macroscopic measurements that can be made in that theory.  And
in our semi-classical state that tracks Minkowski space, we can
add the extra expectation of Lorentz invariance.  This has the
important consequence that any frame-dependent measurements
considered macroscopic with respect to one frame $F$ must be
considered macroscopic with respect to frames that are boosted
with respect to $F$.

As noted in appendix \ref{a:semi}, no such measurements should
alter the state of the system in any detectable way, and so they
should not affect each other -- there must come a point at which
the presence or absence of wave-function collapse ceases to be
physically significant, and this point comes after the
semi-classical scale of the given system is reached.  By analogy,
we can define an approximate measurement of one component of the
angular momentum of a baseball, that does not significantly alter
the state; likewise, the measurement can be made in other frames
rotated with respect to the first, as many of them as we choose,
without causing detectable changes in the state. (Otherwise, to
repeat a point made in the appendix, the question of whether wave
function collapse occurred when we photographed a spinning ball,
or later when we developed the photograph, would become
physically significant. Such theories are physically
unacceptable.)

This is perhaps a source of the difference between the arguments
presented here, and those of \cite{Rovelli:2002vp}. There, the
non-commutativity of observables corresponding to measurements in
different Lorentz frames is emphasised.  The arguments above
contrast with this. They basically assert that for semi-classical
states, there is a sense in which non-commutativity must become
practically insignificant for some class of approximate
measurements, in order to prevent macroscopic quantum interference
effects. This is the situation in standard theories, and there
seems to be no reason to change the requirement in the case of
quantum gravity.

Following the reasoning of the previous section, all of these
macroscopic observations must correspond to properties of the
underlying histories.  Further investigation shows that only
certain theories can maintain Lorentz invariance under this
condition.

\subsection{A simple example}
\label{s:simple}

A light-cone lattice offers a simple discretisation of 2D
Minkowski space, with a fairly obvious discrete/continuum
correspondence. This example offers some interesting insights.

Let the histories of our theory be examples of such a lattice,
with a scalar field living on it.  More specifically, the
structure is a directed graph on a set of elements $\{e(i,j)\}$
with $i,j\in \Integers$.  The graph edges run between the pairs
$\{e(i,j),e(i+1,j)\}$ and $\{e(i,j),e(i,j+1)\}$.  To each element
is associated a real number $\phi(i,j)$ to represent the scalar
field.  Different values of the field give rise to different
histories.

These histories are not defined in terms of a manifold at all.
They are purely ``abstract graphs".  Also, the labels given to the
elements need not be considered to be part of the structure; they
can be reconstructed from the directed graph edges up to
``discrete translations'' (\ie $i \rightarrow i+ x$, $j
\rightarrow j + y$, $x,y\in \Integers$)\footnote{More properly,
the histories should be considered to be equivalence classes of
the directed graphs under the above relabellings of the
elements.}. The only ``intrinsic" properties of an element are the
graph edges connected to it and value of the field there.

Despite this, a correspondence to 2D Minkowski space can be made
in an obvious way: the elements can be assigned to points in
Minkowski with $u=a i$ and $v=a j$, where $a$ is some length
(which might be imagined to be the Planck length), and $(u,v)$ are
the standard light cone co-ordinates in Minkowski space, in terms
of which the metric is $ds^2=2dudv$. This embedding defines the
manifold approximation: a scalar field $\Phi(u,v)$ on Minkowski
space can be considered as an approximation to $\phi(i,j)$ if a
frame can be found such that $\phi(i,j)=\Phi(ai,aj)$.
Realistically, we would only assume this to some degree of
approximation, and also require that the field $\Phi(u,v)$ was not
``quickly varying'' with respect to the distance $a$.  It can
easily be seen that this correspondence principle is consistent
with the discrete translation invariance of the fundamental
histories.

Note that, as well as defining a field on the Minkowski space, the
fundamental history also defines a frame on the Minkowski space
(in which the $(u,v)$ co-ordiniates are defined), which we will
call the ``lattice frame".  For any single fundamental history,
the lattice frame in the effective Minkowski description is fixed
relative to the field configuration.

How do we begin to talk about Lorentz invariance in this model?
Lorentz invariance only makes sense in the continuum, and so any
Lorentz transformations must be applied at the level of the
effective continuum description.  It will be said that one
effective property $X$ is related to another $X'$ by a Lorentz
transformation $\Lambda$ if every field configuration $\Phi(u,v)$
satisfies property $X$ if and only if $\Lambda \Phi(u,v)$
satisfies $X'$.

The problem with this kind of discretisation is that it is ``not
equally good in all frames".  For a given fundamental history,
some properties that are \textit{decidable} (in the sense of
section \ref{s:eff}) are related by Lorentz transformations to
\textit{undecidable} properties.  For example, a plane wave,
written $\Phi(u,v)=\sin(u/\lambda)$ in the lattice frame, can only
be an approximation to a discrete field when $\lambda>>a$.
Therefore the property ``there is a plane wave of amplitude $A$
and wavelength $\lambda>>a$ with respect to frame $F$, in region
$R$'' is only decidable for a particular fundamental history when
the frame $F$ is sufficiently near to the lattice frame (remember
that the lattice frame is fixed relative to the field
configuration for each history).

\begin{figure*} \centering
\resizebox{4.4in}{2.2in}{\includegraphics{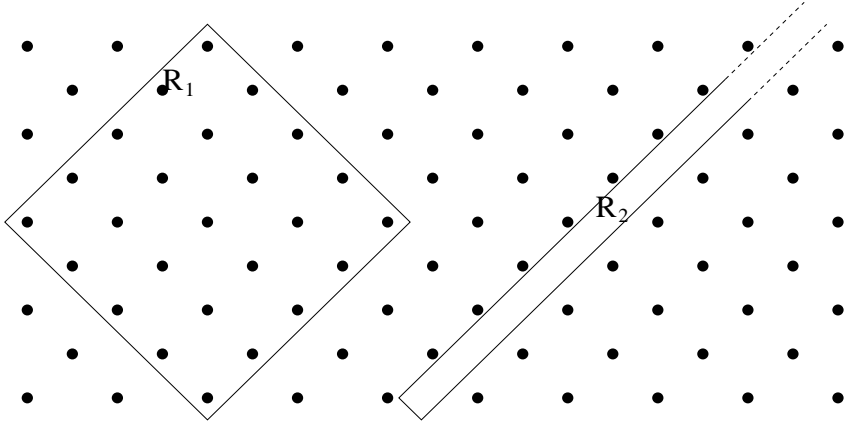}}
\caption{\small{The diagram represents two regions, $R_1$ and
$R_2$, in 2D Minkowski space, into which is embedded a light-cone
lattice as described in section \ref{s:simple}.  $R_1$ is related
to $R_2$ by a boost.  A scalar field defined on the lattice can be
approximated by a field on the Minkowski space.  But it is hard to
imagine how all the properties of the field decidable in $R_1$
could also be decidable in $R_2$, as $R_2$ contains no embedded
lattice points.  With sufficient boosts, similar situations can be
constructed with any finite lattice density and any size of
region, leading to the presence of macroscopic properties of the
history that are decidable in one frame but not another.
 }\label{f:regions}} \end{figure*}

Crucially, there is nothing to stop these properties from being
semi-classical, macroscopic properties.  Even on a very fine
lattice, an extreme enough boost will take a wave of arbitrarily
long wavelength to one with ``sub-lattice'' wavelength.  See
figure \ref{f:regions} for a further example.

\subsection{Lorentz violation, from histories to observations}

This is one example from a class of such discretisation schemes.
A condition on such schemes is now presented that encapsulates
the problem.

\begin{condition}
 \label{c:lv}
 The set of histories of the theory that have Minkowski space
(plus matter) as an effective description is non-empty. Consider
some region $R_1$ in Minkowski and let $X_1$ be a macroscopic
effective property of that region, and also consider another
region $R_2$ related by a boost transformation (with some
sufficiently large boost factor) to $R_1$.  Let $X_2$ be the
property of $R_2$ related to property $X_1$ of $R_1$ by that
boost.  Then for some $X_1$, either $X_1$ or $X_2$ is
undecidable, for all histories. \end{condition}

In the lattice example above, this condition is no doubt
satisfied.  We will now consider a hypothetical realistic model
in which the above condition is also satisfied.  Does this lead
to Lorentz violation?  It might be imagined that the ``lattice
frame'' could vary over the histories in the measured set,
somehow ameliorating the frame dependence of the individual
histories.  This is now shown to be impossible:  the above
condition leads to in-principle observable Lorentz violation.

What if we were to attempt to measure for both properties $X_1$
and $X_2$?  We need the probability associated with the measured
set $\Gamma(X_1) \cap \Gamma(X_2)$.  Unfortunately, either $X_1$
or $X_2$ is undecidable in every history.  This means that the
set $\Gamma(X_1)$ is disjoint from the set $\Gamma(X_2)$, and so
$\Gamma(X_1) \cap \Gamma(X_2) = \emptyset$.  This pair of
macroscopic properties is undecidable in all histories, and hence
always false; it is unphysical.  If there are many easily
measurable pairs like $X_1$ and $X_2$, this is absurd.  The only
solution is that only one of the properties can be considered a
physical, measurable quantity, and so, since one is related to
the other by a boost, Lorentz symmetry is violated.  The
conclusion is that \textit{This kind of Lorentz violation in the
individual histories necessarily leads to Lorentz violation at
the level of macroscopic observables}.

In this argument, no scale has been put on the Lorentz violation
of the individual histories as expressed in condition \ref{c:lv},
hence the vagueness of the phrase ``with some sufficiently large
boost factor''.  But we would expect that Planck scale
discreteness implies Lorentz violation at a related scale, as it
does in the case of the lattice.  A quantification of the effect
would depend on the range of boosts over which a certain property
was well approximated in the individual histories.  No attempt to
quantify this violation for any theory is made here.

In the specific semi-classical situation being described, some of
the symmetries of GR remain; general covariance has been broken to
the Poincar\'e invariance of Minkowski space. How then do we fix
the positions of regions $R_1$ and $R_2$ in all relevant
histories? The positions of the regions have no meaning \textit{a
priori}. They must be determined with reference to something else,
by the properties of other fields in the theory.  Here it is
assumed that this can be done, the reason being that, if it could
not be, then the theory would have failed in any case. It must be
possible to identify in all relevant histories, for example, the
world-line of the planet earth.  With this done, regions could be
identified with reference to it and the above conclusion drawn. In
the lattice example, the world-line of the planet earth would be
described by the configuration of the fields on the lattice, and
would necessarily be identifiable across all histories in the
measured set.  The details of this scheme to fix the positions of
$R_1$ and $R_2$ would not impact the argument that no single
lattice-like history can accurately represent our two macroscopic
properties $X_1$ and $X_2$ (which, it should be noted, may be
defined in overlapping regions, and/or be supplemented with many
additional, similar properties).

\section{Consequences for discrete quantum gravity}
 \label{s:spinfoams}

In order to reach the conclusion, we must assume condition
\ref{c:lv}, a condition arguably true for lattice-like discrete
structures.  The question now becomes: which discrete structures
proposed in the various approaches to quantum gravity satisfy the
condition, and which evade it?

A discretisation scheme along the lines of a regular lattice
would satisfy condition \ref{c:lv}, but that is not a
particularly interesting case; this is not the basis of any
popular attempt to quantise gravity.  In order to fully address
the issue in a given approach, a method of assigning continuum
approximations to the discrete underlying structure would have to
be specified: a map from the individual histories to Lorentzian
manifolds.

In causal set theory, the correspondence between the discrete
underlying structures and the continuum manifold was early on made
explicit, and these discrete structures have been shown to be
Lorentz invariant in a strong sense \cite{Bombelli:2006}, evading
condition \ref{c:lv}. In particular, in contrast to the light-cone
lattice, a scalar field dynamics on a fixed causal set background
can be defined which does not violate Lorentz symmetry
\cite{Henson:2006}.

The situation for spin-foam models is less clear, as here a
correspondence between continuum manifolds and the underlying
histories has not been made completely explicit.  However, some
clues are available.  An analogous situation has been considered
in loop quantum gravity, where discrete spatial configurations
must be approximated by continuous Euclidean manifolds. In
\cite{Bombelli:2004si}, a correspondence is developed for the
simple but suggestive case of unlabelled graphs.

A graph can be associated to a manifold by a random process.
First, a locally finite set of points is randomly selected from
the manifold by a Poisson process (the resulting set of points
being called a ``sprinkling''). To this sprinkling, a graph is
associated via a Voronoi procedure \cite{Bombelli:2004si}.  A
graph is said to be approximated by a manifold if it could have
arisen in this way from a sprinkling of that manifold, with
relatively high probability.  This in-principle correspondence can
be reversed to order to recover geometrical information (\eg on
area, volume and dimension) directly from a given graph.  To allow
for quantum behaviour at short scales, the discrete/continuum
correspondence could be applied only on large scales, by using
some coarse-graining of the fundamental graph
\cite{Bombelli:2004si}.

In this way many desirable features are recovered, for example a
proportionality between the area of a surface of co-dimension 1,
and the number of edges crossing it (something desirable from the
standpoint of loop quantum gravity), and also a proportionality
between the number of vertices in a region and its volume.  Thus
it is plausible that the scheme could be extended to spin
networks. Since these are nothing but spatial slices of
spin-foams, a similar scheme might be sought in that case, keeping
these necessary proportionalities.

However, a na\"ive application of such a scheme to spin-foams
could not be Lorentz invariant.  It has been proven that no finite
valency graph can be associated to a sprinkling of Minkowski space
in a way that respects Lorentz symmetry \cite{Bombelli:2006}, and
each spin-foam contains such a graph. Thus there would always be a
preferred frame associated to each history like the lattice frame
of section \ref{s:simple}, and it is reasonable to assume that
condition \ref{c:lv} would be violated in this case. Without the
randomness of sprinkling, the correspondence might have to rely on
some more regular embedding of vertices into manifolds, something
even more likely to produce Lorentz violations, as it occurs for
regular lattices.

It could be argued that the correspondence principle will be more
subtle, not relying on attempts to embed vertices into manifolds.
But this would seem unlikely in the light of the successes of the
scheme for spatial configurations mentioned above.  Moreover, it
is hard to imagine how to avoid the concept of embedding, at least
at a coarse-grained level, if the vertices of the discrete
structure are to correspond in any way to spacetime points or
``elementary regions''.

But there is another objection to this line of argument. The
Lorentz violation on a lattice is a result of the discreteness.
The finer the lattice, the higher the boosts at which it can
successfully represent macroscopic properties.  But for any
lattice there will come a boost at which these properties cease to
be decidable. Condition \ref{c:lv} requires that there is a boost
factor sufficient to bring about this situation in all histories
that correspond to Minkowski.  This clearly implies that there is
an upper bound on the boost necessary for each individual history.
In the case of spin-foams, this might be questioned. Could there
be no upper bound to how ``fine'' an spin-foam can become, in the
appropriate sense?

The form of the sum-over-triangulations in spin-foam quantum
gravity has not been fixed, and so only more speculation can be
offered on this point. An embedding scheme that put no upper bound
on the number of vertices per unit of 4-volume could possibly
evade condition \ref{c:lv} on these grounds.  It has even been
suggested that a continuum limit may have be taken to get
consistent results from spin-foam models \cite{Baez:2002aw}, in
which case the fundamental histories would not be discrete at all.
But these alternatives seem unlikely to be consistent with the
correspondence of continuum 3-volume and area with properties of
the slices of the underlying spin-foam. Only the establishment of
a full discrete/continuum correspondence principle could
unambiguously answer this question. Even so, it is interesting to
note that for such an evasion to be possible, Planck scale
discreteness would be sacrificed.  The import of this observation
depends upon the value put on Planck scale discreteness as a
desirable aspect of a candidate quantum gravity theory
\footnote{see \eg \cite{Henson:2006kf} for the attitude towards
discreteness taken in the causal set program.}.

\subsection{Further discussion}

Apart from the possible applications of condition \ref{c:lv} to
spin-foams, there are other possible objections to the argument
laid out here that should be mentioned.  For example, some
theories may not satisfy condition \ref{c:soh}; certain approaches
to quantum gravity might not have any SOH formulation in their
final forms. But this seems unlikely to affect the arguments on
spin-foams, especially when the goal of the program is expressed
as finding a path-integral theory for quantum gravity (as, for
example, in \cite{Oriti:2001qu}).  Any theory that did not satisfy
condition \ref{c:soh} would differ considerably from standard
quantum mechanics; in that case many questions would arise
concerning the consistency of the theory and the interpretational
framework to be used.

Secondly, there is some lack of precision in the idea of
``effective descriptions'' here, but it would be difficult to
argue that a lattice-like structure does in fact have enough
structure to represent waves in all frames.  Similarly, although
all the ``macroscopic properties'' of quantum gravity are not
described, they must include the well-observed macroscopic
properties that are present in today's successful theories.

In the argument, it was assumed that the quantum gravity theory
would have semi-classical solutions corresponding to Minkowski
space, and this was used in subsequent arguments.  This might not
be possible in principle, and it is not a realistic scenario. But
it should be possible to generalise to a situation in which
Minkowski was only a good approximation in a region, without
altering the main points of the argument.  The failure of
lattices, and similar structures, to represent macroscopic
properties at very high boosts, is not a subtle effect but a very
marked one, and so it is unlikely to be removed by small
corrections to the picture given above.

Other objections might be found in the practicality of finding and
measuring pairs of events such as $X_1$ and $X_2$, but one would
have to explain why ``an electromagnetic wave with given frequency
and amplitude, travelling through a region $R_1$'' is not a good
property to use for $X_1$, if it is accepted that properties of
this form are measured for in tests of Lorentz invariance. The
biggest problem is that there is no concrete prediction for the
scales at which Lorentz violation would be seen, and this leaves
the door open to Lorentz violation at unmeasurably high boosts.
Thus the argument presented here should be seen as only one step
in the ongoing discussion of Lorentz violation.

\section{Conclusion}

The argument set out above works from the sum-over-histories view
of quantum mechanics.  Firstly, it is argued that all macroscopic
observations correspond, in a fairly direct way, to properties of
histories in the measured set of histories.  If a semi-classical
state is invariant under some symmetry, then the class of
observables that can be considered to be macroscopic will also be
invariant under this symmetry. In a semi-classical state of
quantum gravity that tracks Minkowski space, we expect to be able
to make macroscopic measurements in any frame.  Above, a condition
has been set down under which this is impossible, as there are not
enough properties in the underlying histories to represent these
observations.   Some discussion has been given of the relevance to
the spin-foam and causal set quantum gravity programs.

One purpose of the article was to call attention to the necessity
of an in-principle correspondence between fundamental histories in
a quantum gravity theory, and Lorentzian manifolds which
approximate to them.  In the case of spin-foams, even establishing
some broad features of this correspondence would enable some
conclusions to be drawn from the arguments above, and would
doubtless be useful elsewhere.

Finally, no discussion of doubly special relativity has been made
here, and it is possible that this allows some compromise between
this line of reasoning and those put forward elsewhere, for
example in \cite{Livine:2004xy}.  In conclusion, the arguments
given may put the spin-foam program in the happy situation of
predicting effects that are almost within reach of observation, or
conversely, the principle of Lorentz invariance could be used to
decide amongst various spin-foam models.  But these applications
are dependent on further developments in the ongoing discussion of
Lorentz violation in quantum gravity.

The author is grateful to Jeremy Butterfield, Adrian Kent, Fay
Dowker, Rafael Sorkin and Sumati Surya for correspondence and
discussions of this work, and to the organisers and participants
in the Loops '05 conference, many of whom contributed to the
debate on this issue.  The author was supported by DARPA grant
F49620-02-C-0010R at UCSD, where some work on the article was
carried out.

\appendix

\section{Appendix: Properties individual histories in the SOH}

\subsection{The significance of measurements in one basis}
 \label{a:onebasis}

In section \ref{s:SOH} of the main text, it was shown how
measurements in the histories basis correspond to properties of
histories in the measured set. It was also briefly argued that all
macroscopic observations can be expressed as measurements in this
one basis, \eg the position bases.  This discussion is expanded on
here.

Above, it was claimed that the approximate momentum of a
non-relativistic particle in a semi-classical state can be
determined from a series of position measurements.  It is
interesting to note that this is true even though the
sum-over-histories is dominated by nowhere-differentiable paths,
and so the classical meaning of momentum is not immediately
applicable.  It only comes back into relevance at large scales. In
other words, at scales above the spread of the wave-function, a
typical path approximates to the classical one, and the momentum
can be approximately defined at those scales in the a way that
corresponds to the classical definition.

This would apply to any system, not just the position and momentum
of a particle.  In some approaches to quantum gravity, the history
is a path in superspace (\ie a sequence of 3 manifolds).  It
should in principle be possible to set up experimental apparatuses
in order to measure the intrinsic geometry of a spatial
hypersurface at different times (time, and the foliation
considered, being decided by the experimental apparatus).  In an
informal sense this happens to us all every day -- we are equipped
with an obvious time parameter, and we observe approximately flat
space.  It has been conjectured by Wheeler \cite{Wheeler:1964}
that a sequence of 3D Riemannian manifolds parameterised by time
should be enough to uniquely reconstruct a 4D solution to the
equations of motion of GR. Therefore, we are able, by making a
series of approximate measurements in one basis (that of the
intrinsic geometry of a hypersurface), to reconstruct the
conjugate variables (in this case, the extrinsic geometry).  Our
eventual quantum theory of spacetime should allow this to be
possible for semi-classical states, such as the one which we
observe. This requirement would not be altered if Lorentzian
manifolds were taken to be an approximation to some other
(possibly discrete) structure.

\subsection{Feynman's argument}
\label{a:Feynman}

Although it is stronger than necessary for the purposes of the
arguments of sections \ref{s:LI} and \ref{s:spinfoams}, for
completeness it is interesting to see how measurements in other
bases were dealt with by Feynman:

\begin{quote} The [SOH measurement] postulate is limited to
defining the results of position measurements.  It does not say
what must be done to define the result of a momentum measurement,
for example.  This is not a real limitation however, because in
principle the measurement of momentum of one particle can be
performed in terms of position measurements of other particles,
e.g. meter indicators.  Thus, an analysis of such an experiment
will determine what it is about the first particle which
determines its momentum. \end{quote}

Here Feynman is arguing that the amount of information available
from measurements \textit{in this one basis} is good enough to
describe the results of all experiments.  It is assumed that the
correct correlation can indeed be set up between the meter and the
quantum system, in a measurement situation.  But the possibility
of such correlations is indeed necessary if we are to be able to
make any consistent connection with experiment, as noted in
\cite{VonNeumann:1955vi}.

This is a method of casting all measured properties as properties
of histories in the measured set.  But this is not directly useful
in the present discussion, since the property of the histories
corresponding to, say, the momentum of a particle in a measurement
situation, will not be immediately recognisable. This method
offers no easy connection between simple properties of histories
and observations.  But with macroscopic observations of systems in
semi-classical states we have the more direct given at the end of section
\ref{s:SOH} -- there is no need to couple a pointer to a
baseball in order to know its momentum.

\subsection{The necessity of semi-classical behaviour}
 \label{a:semi}

Here it has been assumed that there exist semi-classical states,
with small uncertainty (relative to the required experimental
accuracy) in all the macroscopic variables that we might want to
measure. In particular, one must be able to approximately measure
macroscopic quantities without noticeably affecting the state.
This property of standard quantum mechanics must also be true for
any future quantum theory of gravity.  It is not an incidental
property of current quantum theories, but is essential for the
theoretical framework to make sense.

Von Neumann stressed the requirement \cite{VonNeumann:1955vi} that
we have the freedom the change the classical/quantum boundary at
which wave-function collapse occurs.  In an experimental set-up
there is a division of the universe into quantum system, and
classical apparatus and environment.  This division must be made
so that none of the \textit{microscopic} degrees of freedom of
interest are taken to be classical; but any choice of division
that accords with this rule-of-thumb is valid, and the choice
should not be physically significant. To maintain this principle,
it must be possible to (approximately) measure
\textit{macroscopic} properties, without significantly altering
the state.  Our theory would be faulty if an approximate
measurement of the position of a rock necessarily affected the
results of subsequent approximate measurements. In such a case of
``macroscopic complementarity'', the imposition of wave-function
collapse has detectable consequences, opening a Pandora's box of
conceptual problems that are kept at bay in standard quantum
theories.

A lack of semi-classical states, with small quantum fluctuations
in all the quantities that we expect to be behave approximately
classically, would create such problems.  The allowable
fluctuations must be decided with respect to experimental
constraints and expectations stemming from these constraints, as
noted in \cite{Ashtekar:2005dm}.  This allows such results as that
of \cite{Ashtekar:1996yk}, in which a symmetry reduced
gravitational system is quantised, and relatively large
fluctuations in the metric are found to be possible, even in
coherent states. The situation described there is far from being
one that we have experimental access to, and so ``reasonable
expectations'' for the magnitude of quantum fluctuations should
be, in this case, very loose.  In the light of this, the result
need not be viewed as a failure of semi-classicality, but only as
a situation in which the meaning of semi-classicality is unusual.
But we are not at all far from experiment when discussing
approximately flat spacetime, and so the constraints here are much
tighter, especially if one of our expectations is the survival of
Lorentz invariance.

\subsection{Summary}

The conclusion is that \textit{each individual history in the
measured set must have all of the properties that we observe}.
Extracting these observations when given such a history may not be
easy, but it must in principle be possible.  For semi-classical
quantities, the relationship to properties of histories should be
direct, allowing us to proceed as we do classically, albeit on a
suitably course-grained scale at which the classical quantities
make sense.  This is the case in standard quantum theories. In the
original example, a typical path of a Schr\"odinger particle might
be highly fractal on small scales, but in semiclassical situations
it will approximate the classical solution with an acceptable
degree of accuracy.  Any series of position measurements of a
macroscopic object could be said to be an informal verification of
this aspect of the theory.

Conversely, there are no measured properties that belong only to
sets of histories, and not to its members. Such properties fall
into two other categories: (a) Properties in other bases that are
never measured, and so are never correlated to a ``meter''
property in the way that Feynman requires; (b) Mathematical
attributes that do not represent the outcomes of any real or
imagined experiment, for example, the property of a set of
histories decohering with another set of histories (see \eg
\cite{Hartle:1992as}).

Feynman's remark that all observations can be handled in one
basis is little known, and sometimes explicitly contradicted, in
modern work on theoretical physics.  The point seems to be raised
only in work on the measurement problem and its various proposed
solutions, \eg
\cite{Bell:1987,Sorkin:1995nj,Kent:1997bc}\footnote{An argument
is made in \cite{Sorkin:1995nj} and also \cite{Kent:1997bc} to
the effect that it is possible to regard one history from the
measured set as the ``real one'', since it contains all measured
properties.}. However, this picture of quantum mechanics is not
only of interest from this perspective.  It impacts on an
important current issue in quantum gravity: exploring the
properties of semi-classical states.

\bibliographystyle{h-physrev3}
\bibliography{refs}
\end{document}